\documentclass[reprint, amsmath, amssymb, aps, prb]{revtex4-1}

\usepackage[]{graphicx}      
\usepackage{bm}             	
\usepackage{times}			
\usepackage{tabularx}
\usepackage{color}
\usepackage{array}
\usepackage{float}
\usepackage{epstopdf}

\newcolumntype{P}[1]{>{\centering\arraybackslash}p{#1}} 
\newcolumntype{M}[1]{>{\centering\arraybackslash}m{#1}} 

\newcommand{\IEF}{Institut d'Electronique Fondamentale, CNRS, Univ. Paris-Sud, Universit\'{e} Paris-Saclay, 91405 ORSAY, FR}
\newcommand{\SAMSUNG}{SAMSUNG Electronics Corporation, 601 McCarthy Blvd Milpitas, CA 95035, USA}

\begin{document}


\title{Spin-wave thermal population as temperature probe in Magnetic Tunnel Junctions}

\author{A. Le Goff}
\email{adrien.le-goff@u-psud.fr}
\affiliation{\IEF}



\author{V. Nikitin}
\affiliation{\SAMSUNG}
\author{T. Devolder} 
\affiliation{\IEF}

\date{\today}                                           
%
%
\begin{abstract}
We study whether a direct measurement of the absolute temperature of a Magnetic Tunnel Junction (MTJ) can be performed using the high frequency electrical noise that it delivers under a finite voltage bias. Our method includes quasi-static hysteresis loop measurements of the MTJ, together with the field-dependence of its spin wave noise spectra. We rely on an analytical modeling of the spectra by assuming independent fluctuations of the different sub-systems of the tunnel junction that are described as macrospin fluctuators. We illustrate our method on perpendicularly magnetized MgO-based MTJs patterned in $50\times100$ nm$^2$ nanopillars. We apply hard axis (in-plane) fields to let the magnetic thermal fluctuations yield finite conductance fluctuations of the MTJ. Instead of the free layer fluctuations that are observed to be affected by both spin-torque and temperature, we use the magnetization fluctuations of the sole reference layers. Their much stronger anisotropy and their much heavier damping render them essentially immune to spin-torque. We illustrate our method by determining current-induced heating of the perpendicularly magnetized tunnel junction at voltages similar to those used in spin-torque memory applications. The absolute temperature can be deduced with a precision of $\pm 60$~K and we can exclude any substantial heating at the spin-torque switching voltage.
\end{abstract}


\maketitle

%
%

\section{Introduction}

A strong research effort is currently done within the spintronics community to obtain ever improved magnetic properties \cite{sbiaa_reduction_2011, jan_high_2012}for optimal utilization in integrated devices like sensors and memories \cite{thomas_perpendicular_2014} which often rely on Magnetic Tunnel Junctions (MTJ). Often overlooked, the evolution of the magnetic properties with temperature is crucial for achieving high performance MTJ. Indeed, most of the basic properties of an MTJ are temperature dependent \cite{xiao_temperature_2014} like the saturation magnetization $M_\textrm{S}$, the anisotropy field $H_\textrm{k}$ and the spin-polarization of the ferromagnetic electrodes and thus the spin-polarized transport properties. Besides, thermal fluctuations also impact the stability of the information in memory devices like spin-torque magnetic random access memories (STT-MRAM). The reliability --and consequently the energy cost-- of magnetization switching is also affected by temperature-induced fluctuations \cite{bedau_ultrafast_2010}. Furthermore, any rise of the temperature can accelerate the material fatigue of the MTJ: this is not only by enhancing the rate of interdiffusion of the materials constituting the MTJ but also by favoring thermally assisted dielectric breakdowns of the oxide barriers. Measuring the temperature of an MTJ in which a current is passing is thus of fundamental importance.

Several methods have been proposed to measure the temperature of MTJs during operation. The measurement of the exchange bias field supplied by the antiferromagnetic layer of an MTJ is one of these methods. Unfortunately it cannot be applied to state-of-the-art perpendicularly magnetized MTJs which are most often not exchanged biased \cite{{xiao_temperature_2014},{papusoi_reversing_2008}, {gapihan_heating_2012}, {sousa_tunneling_2004}}. The other methods to determine the MTJ temperature are indirect: they include the statistical study of the distributions of coercive fields $H_\textrm{C}$ \cite{tsai_intrinsic_2013} or switching currents  $I_\textrm{C}$ \cite{hosotani_effect_2010}. These statistical methods are undermined by their questionable description of the switching phenomena and the influence of the temperature thereon. Advanced models like the numerical solving of the heat diffusion equation using finite element codes \cite{sousa_crossover_2006} have usually large uncertainties because they often rely on thermal conductivities that are loosely known at the nanoscale.

In this paper, we report a direct method to measure the absolute temperature of a tunnel junction under finite voltage. We illustrate the method on STT-MRAM cells biased with hard axis applied fields. We perform spin wave noise spectroscopy (SWNS) by spectrally analyzing the voltage noise that the MTJ delivers to an external circuit. The frequency of each spin wave mode has a specific field dependence that is used to assign each mode to a specific magnetic sub-system host within the MTJ. We select the main mode of the reference system of the MTJ; the population of this mode reflects the MTJ temperature that can be deduced using a transparent analytical model.  \\
The paper is organized as follows. The samples and their conventional properties are described in section \ref{static}. The noise spectroscopy experiment is depicted in section \ref{exp}. The relation between the magnetization noise and the electrical noise is then modeled (¤\ref{model}) and discussed on the practical case of an STT-MRAM cell (¤\ref{discu}). We obtain a limited evolution of the device temperature for applied voltages below 700 mV and a gradual heating for higher applied voltages, till the junction reaches $420\pm 60$~K for a bias $U_\textrm{DC}$ = 0.9 V) just before the dielectric breakdown that happens at 1~V.

 
\section{Hysteresis and quasi-static properties \label{static}}

Our samples have the following structure: Substrate / Bottom Electrode (BE) / Hard Layer (HL) / Ru / Reference Layer (REF, 3 nm) / MgO / Free layer (FL) / MgO (capping layer). They are patterned into $50\times100$~nm$^{2}$ quasi-rectangular pillars . The HL subsystem is a [Co/Pt]$_{\times n}$ high anisotropy multilayer. The REF and FL are the polarizing and storing FeCoB layers respectively. All layers have a magnetization perpendicular to the sample plane at remanence. 

The static measurements consist of resistance versus field loops undertaken on the pillars. Typical minor and major hysteresis loops with easy axis fields are displayed in Fig.~1(a) and (b). The free layer minor loop indicates the very accurate compensation of the stray field of the reference Synthetic AntiFerromagnet (SAF) composed of the REF and HL. A field of 1 Tesla results also in a minor loop, in which only the magnetizations of the FL and the REF of the SAF are manipulated : the loop resembles that of a spin-valve [Fig.~1(a)]. A field higher than 1.45 T is required to switch also the hardest layer (HL) of the SAF [Fig. 1(b)]. When saturating the three layers, the next remanence is always in a high resistance (antiparallel) configuration. The tunnel magneto-resistance ratio reached up to 110 $\%$ at small bias voltage.

Resistance versus in-plane field (i.e. hard axis field) were also performed [Fig.~1(c)]. In such loops, the inflection of the slope indicates the saturation of the softest layer along the in-plane direction at the effective anisotropy field value $H^\mathrm{{eff, FL}}_\mathrm{k}$ of the FL. A saturation field $\mu_\textrm {0}$ $H_\mathrm{k}^\mathrm{{eff, FL}}$ of 500 $\pm$ 25 mT is found at low voltage, and this decreases in a  parabolic manner at higher voltages. 


\begin{figure}

\includegraphics[width=9cm]{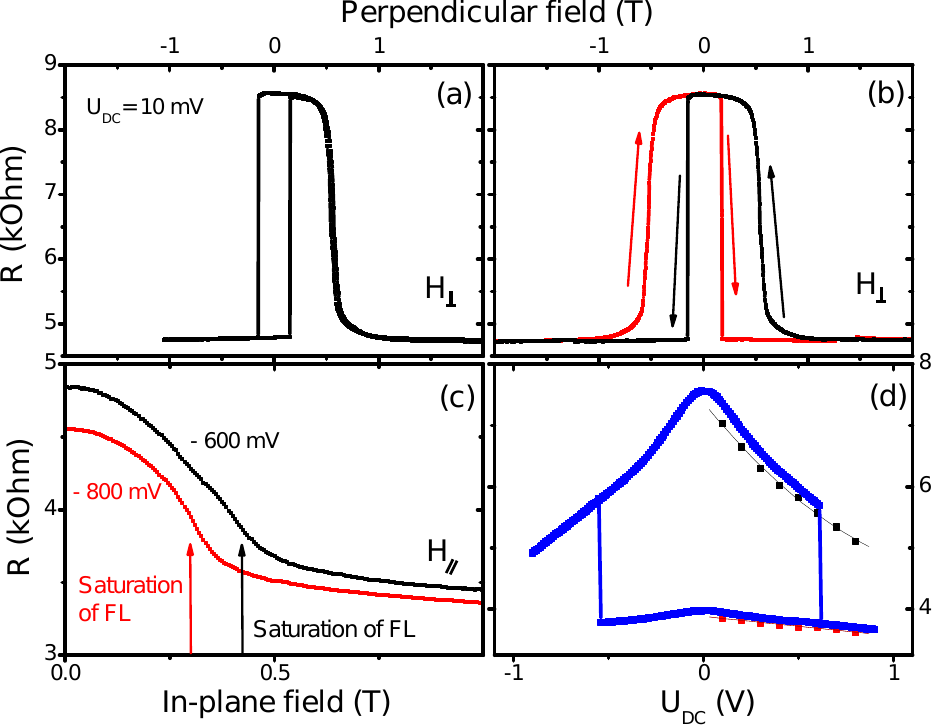}
\caption{Quasi-static hysteresis loops of the tunnel junction. Panels (a) and (b): resistance versus easy-axis field loops. a) Minor loop (free layer and softest layer of the reference SAF). (b) Major hysteresis loop. The arrows denote the field sweeping direction. Panel (c): resistance versus hard-axis field for two different voltages applied across the MTJ. The arrows point at the inflection of the loop slope which is indicative of the saturation of the FL. (d) Resistance versus applied voltage bias at zero field (STT hysteresis loop, blue curve). The P and AP states can be stabilized at higher voltages using large perpendicular fields. The black and red symbols illustrate this fact, and they were collected during the noise spectroscopy measurements (§III).}
\end{figure} 

The STT hysteresis loop in vanishing applied field is displayed in Fig.~1(d). It exhibits the classical house shape, with a much pronounced I-V non linearity in the antiparallel (AP) high resistance state. The switching voltage is approximately 60\% of the breakdown voltage value (1 V), which allows safe long experiments up to 0.9 V. We will see later that we need to know the dependence of the resistances $R_\mathrm{{P}}$ and $R_\mathrm{{AP}}$ of the P and AP states at voltages larger than the switching voltage; we thus have superimposed on the STT loop the voltage dependence of the $R_\mathrm{{P}}$ and $R_\mathrm{{AP}}$ by stabilizing these states using large applied fields.

 \section{Spin Wave Noise Spectroscopy \label{exp}} 
 \subsection{Principe and experimental procedures}

The Spin Wave Noise Spectroscopy relies on TMR (Fig. 2). Indeed, the conductance of the MTJ scales with the cosine between the directions of magnetizations on each side of the tunnel barrier, i.e. at the FL and the REF. Any fluctuation of one of these magnetizations induces a conductance fluctuation (i.e. electronic noise) which can be recorded. The precession angle of the magnetization is a direct measurement of thermal population of spin waves, which will be used to deduce the temperature of the spin wave bath. \\ The first step is to detect the eigenmodes of precession. For this we have followed the experimental procedure detailed in \cite{devolder_spin-torque_2011} which consists in measuring the voltage noise that the MTJ delivers to a low noise broadband amplifier and then to spectrally analyze this noise for variable applied fields and applied voltages.We have applied the field in the in-plane direction (along the short axis of the MTJ) in order to break the circularity of the precession eigenmodes of PMA systems and consequently make these eigenmodes detectable by magneto-resistive techniques [Fig.~2(b)]. 

In past studies \cite{helmer_quantized_2010, devolder_spin-torque_2011}, SWNS was used to deduce the magnetic properties only, hence the mode frequency and linewidth were the sole informations that mattered. Conversely, our present temperature measurement requires the precise determination of the \textit{power} that each spin wave population carries. The frequency-dependent gain of our measurement chain (RF cable assembly, bias tees and low noise amplifiers) was thus calibrated carefully by measuring its scattering matrix using a vector network analyzer and calibration-grade attenuators in the 200 MHz-50 GHz band. When connecting then the device with an RF probe, we measured the noise of the experimental set-up in the absence of the magnetically induced noise. This is done by not biasing the sample (i.e. keeping U$_{DC}$ = 0 across the junction). The noise spectrum so collected on the spectrum analyzer was checked to be consistent with a noise figure always better than NF=5 dB for the ensemble of the measurement chain plus the RF probe. \\
The electrical noise related to magnetic fluctuations is then turned on by biasing the device with a constant voltage U$_{DC}$ $\neq$ 0. After division by the gain of our measurement chain, we subtract the previously measured noise at U$_{DC}$ = 0 to isolate the \textit{magnetic-only} noise as would be induced by the sole magnetization fluctuations.  Note that because of this imperfect subtraction procedure, the unavoidable noise in the measurement of any noise power spectra can lead to slightly negative power densities at some frequencies in the displayed magnetic-only noise power densities [see Fig. 2(c)]. This will be accounted for by the factor $\epsilon$ in the section \ref{confidence}. 
The above procedure was repeated for variable applied fields and variable applied voltages.  Fig. 2(c) shows one example of spectrum obtained for a field of -1T. 
 
\begin{figure}
\includegraphics[width=8.5cm]{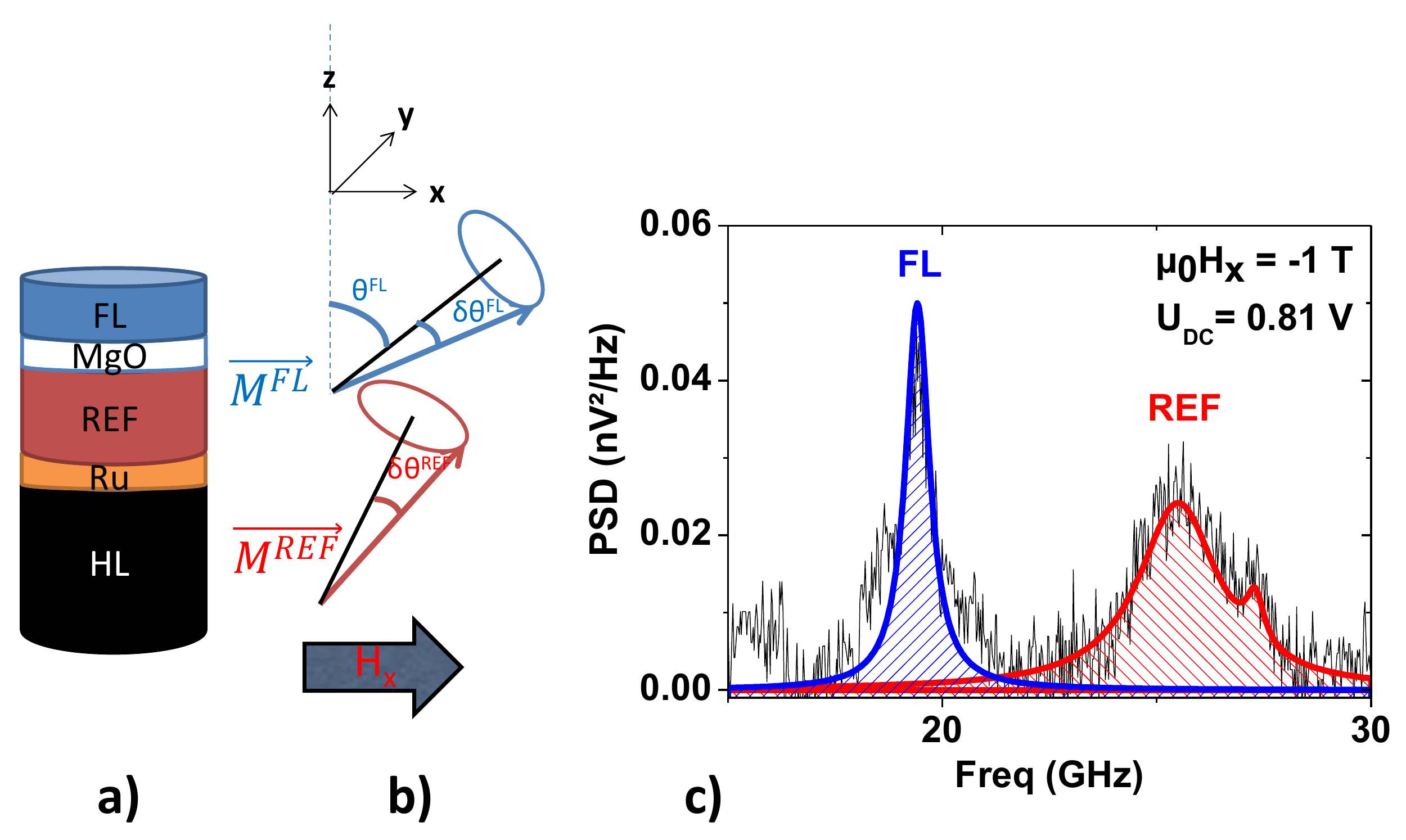}
\caption{Principle of the spin-wave noise spectroscopy experiment. (a) Sketch of the MTJ structure. (b) Definition of the coordinate axes and the angle convention. Sketch of the fluctuations of the free layer magnetization (blue ellipse) and reference layer magnetization (red ellipse). The in-plane field $H_x$ breaks the cylindrical symmetry of the system, resulting in elliptical precessions that make the conductance vary. (c) Magnetic-only noise as induced by the thermally excited precession of free and reference layers through TMR. The shaded Lorentzian curves are used to determine the integrated noise power of the quasi-uniform mode of the FL (blue), the quasi-uniform mode of the REF (red), and the non uniform modes of the FL (red shoulder).}
\end{figure}

\subsection{Layers properties deduced from spin wave spectra}
\subsubsection{Free Layer properties}
 


During SWNS experiments, at least two modes are always perceptible [Fig.~(3a)]. By continuity criteria when varying $H_x$ and $U_{DC}$ we can distinguish them from the additional modes (not shown) revealed at high voltages. 
The frequency versus field curve of the lowest frequency mode has a W-like shape. This shape is typical of hard axis dispersion curves, and the apices of the W indicate the saturation of some magnetization along the field direction above a specific field value. This field value indicates that the W-shaped mode is the quasi-uniform ferromagnetic resonance (FMR) mode of the free layer.
After saturation of the FL magnetization along the in-plane field $H_{x}$ , the dependence of its FMR angular frequency is essentially:
\begin{equation}
\omega_\textrm {FMR} = \gamma_0 \sqrt{H_\textrm {x} (H_\textrm {x} - H_\mathrm{k}^\textrm {eff})} 
\label{FMR_freq_FL}
\end{equation}
where $\gamma_0$ is the gyromagnetic ratio, $H_\mathrm{k}^\textrm {eff} = H_\mathrm{k}-(N_z-N_x) M_\textrm{S} = H_\mathrm{sat}$ is the effective anisotropy field at which the FL saturates, and which takes into account the magneto-crystalline anisotropy field $H_\mathrm{k}$ , the magnetization $M_\textrm{S}$ and the demagnetizing factors $N_{y, z}$ of the free layer. Our convention is $H_\mathrm{k}^\mathrm{eff}>0$ for perpendicular anisotropy. Note that in respect of the essentially null value of the offset field of the FL loop, we consider the FL as not coupled with any other layer. Thus, the effective anisotropy field can be deduced from the fitting of Eq.~(1) using measured values of $\omega_\textrm {FMR}$.

To assign each mode and to confirm the effective anisotropy field values, we also measured the value of resistance of the MTJ during SWNS experiments. As the saturations of the FL and REF layers occur at different fields, a sudden rise of the derivative of the MTJ resistance occurs at the saturation of the FL [see the arrows in Fig. 1(c)] and can be used to check its effective anisotropy field. 
Fig.~3(a) sums up the voltage dependence of the FL saturation fields as obtained using either Eq.~\ref{FMR_freq_FL} or using the inflection point in the resistance versus field loops [Fig. 1(c)] . The two techniques yield similar values. The lower accuracy of the resistance derivative technique is compensated by its operating capability in a wider $U_\mathrm{DC}$ interval. In addition to a slight asymmetry, the FL saturation field decays above 0.5 V with a quadratic trend from a maximum of $H_\mathrm{k}^\textrm {eff, FL}= 550 \pm 60$ mT at 0.1 V to  300 $\pm$ 30 mT at -0.8V. The slight voltage asymmetry is probably a consequence of the spin-torque acting on the FL, while the quadratic decay can either be due to the sole current-induced heating or to its conjunction with STT. This intricacy of the consequences of STT and of current-induced heating makes it difficult to use the FL properties to deduced the sample temperature. 
Let us thus turn our attention to the reference layer.

\subsubsection{Reference layer properties}

The frequency versus field curve of the highest frequency mode has a V-like shape, which indicates that the maximum field ($\approx 1.6$~T) at which we detect this mode is not strong enough to saturate the magnetization of the layer hosting this mode. The mode high frequency and the large saturation field implies that the layer hosting this V-shaped mode has an anisotropy field much higher than the FL. However this V-shaped mode cannot be assigned to the magnetization precession in the sole HL for three reasons. 
(i) The high damping value of Pt-rich hard layer would yield much larger linewidth for this mode.   
(ii) From the minor and major loops, we saw that the saturation field of the HL is above 1.45 T, which implies a zero field FMR frequency above typically 40 GHz for the modes hosted by the HL while we measure only 18 GHz. 
(iii) Furthermore, as the SWNS technique relies on TMR, we are not directly sensitive to the HL because the HL is not in contact with the MgO tunnel barrier. \\
We thus ascribe the V-shaped mode to the REF layer. Although the method reported in this paper does not require to know the exact properties of the HL and its coupling with the REF layer, we have used the frequency versus field dispersion curve to yield estimates of the properties of the REF and HL following standard methods \cite{helmer_quantized_2011}.
In the remainder of the paper, we shall see that it is sufficient to describe the REF layer with a large saturation field of $H^\textrm {eff,~REF}_\textrm {k}$ = 2 T. Also, as in practice only the saturation fields will matter for the temperature measurements, we will lighten our notation to  $H^\mathrm{{}}_\mathrm{sat}$ instead of $H^\mathrm{{eff}}_\mathrm{k}$. 

Finally we mention that in addition to the  two main noise peaks that were assigned to FL and REF quasi-uniform precession mode,  shoulder int he REF peak or a small separated peak are also detected at the largest bias voltages . The linewidth of this shoulder is comparable to that of the quasi-uniform mode of the FL. We shall see here after that this arises most probably from non uniform spin waves within the FL.
 
\begin{figure}
\includegraphics[width=8cm]{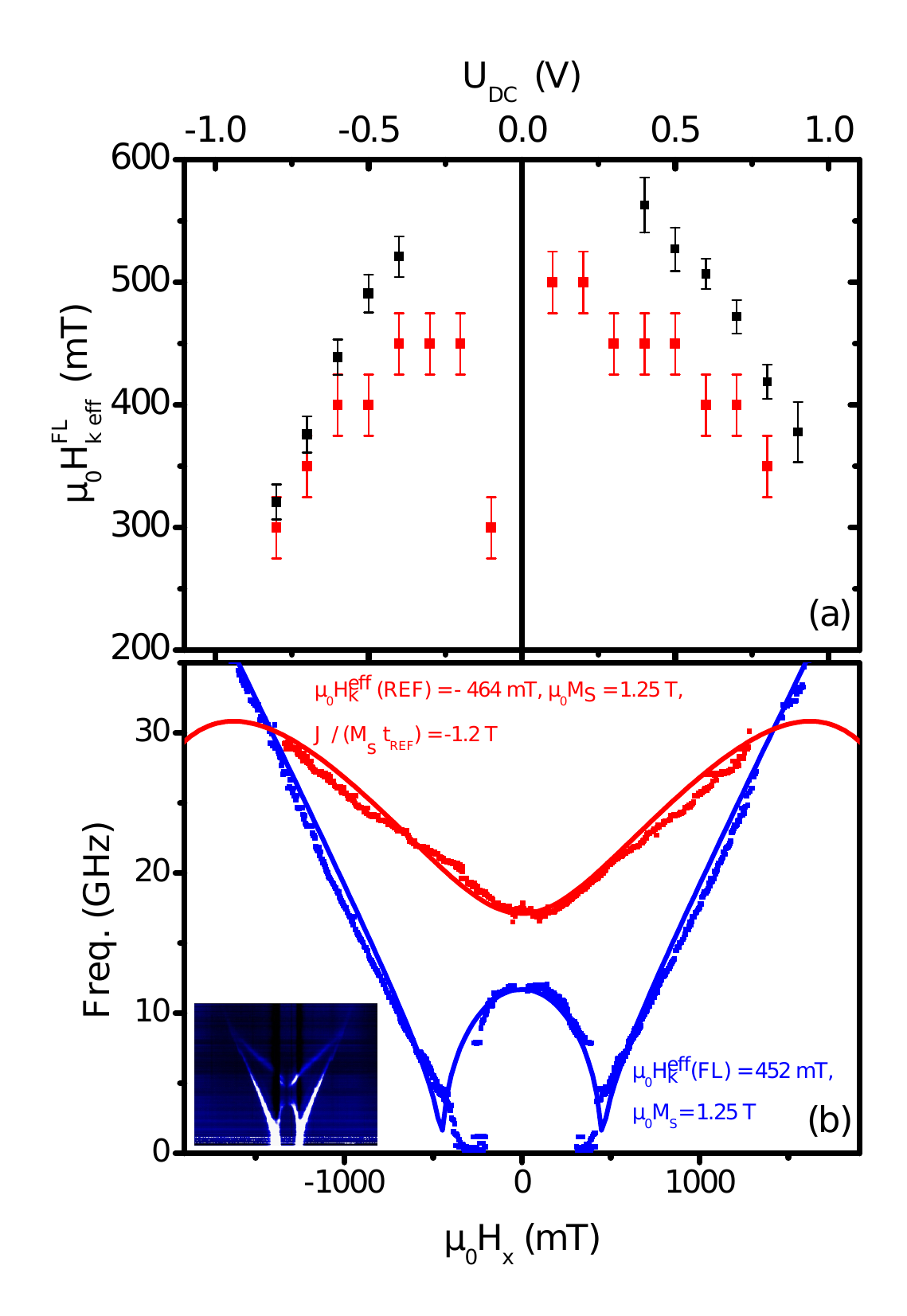}
\caption{(a): Free Layer anisotropy field. The black symbol were extracted using fits to Eq. 1 while the red ones were deduced from the inflection points in the resistance versus in-plane field loops. (b) In-plane field dependence of the frequencies of the quasi-uniform FL mode and quasi-uniform REF layer mode. The symbols are the experimental data and the lines are calculation in a 3 macrospin model following the methods of \cite{devolder_joint_2015} and using the parameters superimposed on this panel and a HL of saturation field of 2 T. The inset shows the representative spectra obtained for an applied voltage of -800 mV. The frequency and field spans in the inset are identical to that of panel (b).}
\end{figure}

\section{From electrical noise to device temperature \label{model}}

Now that each mode was assigned to its hosting layer, let us model the amplitude of the corresponding fluctuations, first in the magnetization domain, then in the conductance domain. 
We start by writing the most simple form of the energy of the layer of interest in the macrospin approximation :
\begin{equation}
 E = -\mu _\textrm {0} M_\textrm{S} H_\textrm {x} \sin \theta \cos \phi  + \frac{1}{2}\mu_\textrm {0 }M_\textrm{S} H_\textrm {sat} \sin^\textrm {2}\theta
\label{energy}
\end{equation} 
where $\theta$ is the angle between the magnetization and the z axis (see Fig.~2b) and $\phi$ is the angle between the magnetization and (x) which is also the field direction. As usual $M_\textrm{S}$ is the magnetization,  and $H_\textrm {sat} $ is the in-plane saturation field of the layer of interest. 
Note that we neglect the difference between the (x) and (y) directions that arises in principle from the non circular shape of the device and the associated demagnetizing contributions. This is justified by the fact that we shall use in-plane fields that exceed the in-plane demagnetizing field by two orders of magnitude.  

The equilibrium is found by letting $\frac{\partial E}{\partial \theta} = 0$ and $\frac{\partial E}{ \partial \phi} = 0$. This gives:
 
$$\left\{
\begin{array}{l}
 \phi_0 = 0 ~[\pi], \ \theta=\pi/2 \textrm{  i. e. saturation when } |H_\mathrm{x}| > H_\textrm{sat}  \\
 \phi_0 = 0 ~[\pi], \ \sin\theta_\textrm {0} =\frac{|H_\textrm {x}| }{H_\textrm {sat}} \\
 \end{array}
 \right.$$
\begin{flushright}
i. e. magnetization tilt when $|H_\textrm {\textrm {x}}| < H_\textrm {\textrm {sat}}$ (see Fig. 4a))
\end{flushright}

\subsection{Fluctuation of the magnetization orientation}

Let us estimate the standard deviation $\delta \theta$ of the magnetization tilt $\theta$ around its equilibrium position  $\{\theta_\textrm {0}, ~\phi_\textrm {0}\}$. For this we expand the energy :

\begin{equation}
E(\theta_\textrm {0}+\delta \theta) - E_\textrm {0} = \frac{1}{2} \mathrm k_\textrm {B} \mathrm T = \frac{1}{2} \frac{\partial^\textrm {2} E}{\partial\theta^\textrm {2}}\delta \theta^\textrm {2}
\label{Taylor}
\end{equation}

which is equivalent to:


\begin{equation}
 \delta \theta = \sqrt[]{\frac{\mathrm k_\textrm {B} \mathrm T  }{V \mu_\textrm {0} M_\textrm{S} }} \sqrt{\frac{H_\textrm {sat}}{(H_\textrm {sat}+H_\textrm {x})(H_\textrm {sat}-H_\textrm {x})}}
\label{DeltaThetaBeforeSaturation}
\end{equation}

$\hspace{5cm} \textrm{~for}\  0<H_\textrm {x}<H_\textrm {sat}$ and,

\begin{equation}
\delta \theta =  \sqrt[]{\frac{\mathrm k_\textrm {B} \mathrm T}{V \mu_\textrm {0} M_\textrm{S} }} \sqrt{\frac{1}{H_\textrm {x}-H_\textrm{sat} }} 
\label{DeltaThetaAfterSaturation}
\end{equation}

$\hspace{5cm} \textrm{~for}\ H_\textrm {x}>H_\textrm {sat} $.


A similar calculation with $\delta \phi$ leads to:

\begin{equation}
\delta \phi = \sqrt[]{\frac{\mathrm k_\textrm {B} \mathrm T}{\partial E^\textrm {2}/\partial \phi^\textrm {2}}}=\sqrt[]{\frac{\mathrm k_\textrm {B} \mathrm T}{V \mu_\textrm {0} M_\textrm{S}}} \sqrt{\frac{H_\textrm {sat}} {{H_\textrm {x}}^2}} 
\end{equation}
$\hspace{5cm} \textrm{~for}\  0<H_\textrm {x}<H_\textrm {sat}$ and,

\begin{equation}
\delta \phi =\sqrt[]{\frac{\mathrm k_\textrm {B} \mathrm T}{V \mu_\textrm {0} M_\textrm{S}}} \sqrt{\frac{1} {{H_\textrm {x}}}} \textrm{~for}\ H_\textrm {x}>H_ \textrm {sat}
\end{equation}

From our experimental data, we can estimate the fluctuation angles [Fig. 4(c)]. Using an MTJ of $50 \times 100 $~nm$^2$ with $t_\textrm {FL}$= 1.8 nm, $t_\textrm {REF}$= 3 nm, $M_\textrm{S}^\textrm {FL}=M_\textrm{S}^\textrm {REF}$=1 MA/m, $\mu_0 H^\textrm {FL}_\textrm {sat}$=0.5 T, and $\mu_0 H^\textrm {REF}_\textrm {sat}$=2 T, we get sizable FL fluctuations reaching $\delta \theta^\textrm {FL}  \approx$ 1.7 degrees at remanence. The harder and thicker REF fluctuates less, with $\delta \theta^\textrm {FL}  \approx$ 0.67 degrees at zero field. Note that the Taylor expansion of the energy (Eq. \ref{Taylor}) progressively looses its relevance when the fluctuation angles diverge and the applied field approaches the saturation fields of either the REF or the FL [Fig. 4(c)]. We will thus not analyze the fluctuations near these fields.

\subsection{Thermal fluctuations of the conductance}

Let us deduce the variation of conductance $\delta G$ induced by  $\delta \theta$ and $\delta \phi$  in the vicinity of $\{ \theta^\textrm {FL}_\textrm {0}, \theta^\textrm {REF}_\textrm {0} \}$. Note that resistance and conductance fluctuations are linked by  $\frac{\delta G}{G} = - \frac{\delta R}{R}$.

The conductance is :
\begin{equation}
G = G_\textrm {0} +\frac{\Delta G_\textrm {max}}{2} \textbf{m}^\textrm {\textbf{FL}}.\textbf{m}^\textrm {\textbf{REF}}
\end{equation}

where $G_0 \approx 1.88\times10^{-4}$~S is the conductance when the two layers have orthogonal magnetizations and $\Delta G_\textrm {max} \approx 1.25\times10^{-4}$~S the tunnel magneto-conductance.

We do the approximation that $\delta \theta$ and $\delta \phi$ are independent fluctuations, and write
\begin{equation}
\delta G\approx G(\theta^\textrm {FL} +\delta\theta, \phi^\textrm {FL}+\delta\phi, \theta^\textrm {REF}, \phi^\textrm {REF})-G_\textrm {0}
\end{equation}


After some algebra, we get that the conductance fluctuation is simply:

\begin{equation}
\delta G = \frac{\Delta G_\textrm {max}}{2} \sin(\theta^\textrm {REF}-\theta^\textrm {FL})  \ \delta\theta\
\label{deltaG_formule_generale}
\end{equation}
The conductance fluctuation is only function of the fluctuation of the magnetization tilts $\delta \theta$ and the angle between REF and FL magnetizations as taken into account in the \textit{sensitivity function} $\sin(\theta^\textrm {REF}-\theta^\textrm {FL})$. This sensitivity function [Fig. 4(b)] is zero when the magnetizations are collinear ($H_\textrm {x}$= 0 T or $H_\textrm {x} \geq H^\textrm {REF}_\textrm {sat}$) and reach a maximum when FL is saturated along $x$ while the REF is not ($H_\textrm {x} = H^\textrm {FL}_\textrm {sat}$).

\subsubsection{Conductance noise induced by Free Layer fluctuations}

For a start, we consider only one fluctuating moment (the Free Layer) in the presence of a static reference layer. We have to distinguish 3 field intervals [Fig. 4(a)] depending on the relative states of the FL and REF magnetizations: \\

\textbf{Case A:} At low fields the free and reference layers have both a tilted magnetization, i.e. $\theta^\textrm {REF}_\textrm {0} < \pi/2$ and $\theta^\textrm {FL}_\textrm {0} < \pi/2$. The conductance fluctuations are :

\begin{equation}
\delta G_\textrm {\textrm{dual tilt}}^\textrm {\textrm{FL~fluctuating}} =  \frac{\Delta G_\textrm {max}}{2} 
\sin (\theta^\textrm {REF}_\textrm {0}-\theta^\textrm {FL}_\textrm {0})~ \beta^\textrm {FL} \gamma^\textrm {FL}
\label{FL_dual_conductance}
\end{equation}

\begin{footnotesize}
\begin{flushright}:
where $\beta^\textrm {FL}=\sqrt[]{\frac{\mathrm k_\textrm {B} \mathrm T  }{\mu_\textrm {0} M_\textrm {S}^\textrm {FL} \textrm V_\textrm {\textrm FL} }}$ \\
and $\gamma^\textrm {FL}= \sqrt{\frac{H^\textrm {FL}_\textrm {sat}}{(H^\textrm {FL}_\textrm {sat}+H_\textrm {x})(H^\textrm {FL}_\textrm {sat}-H_\textrm {x})}}$\\
\end{flushright} 
\end{footnotesize}


\begin{figure*}
\includegraphics[width=0.8 \paperwidth]{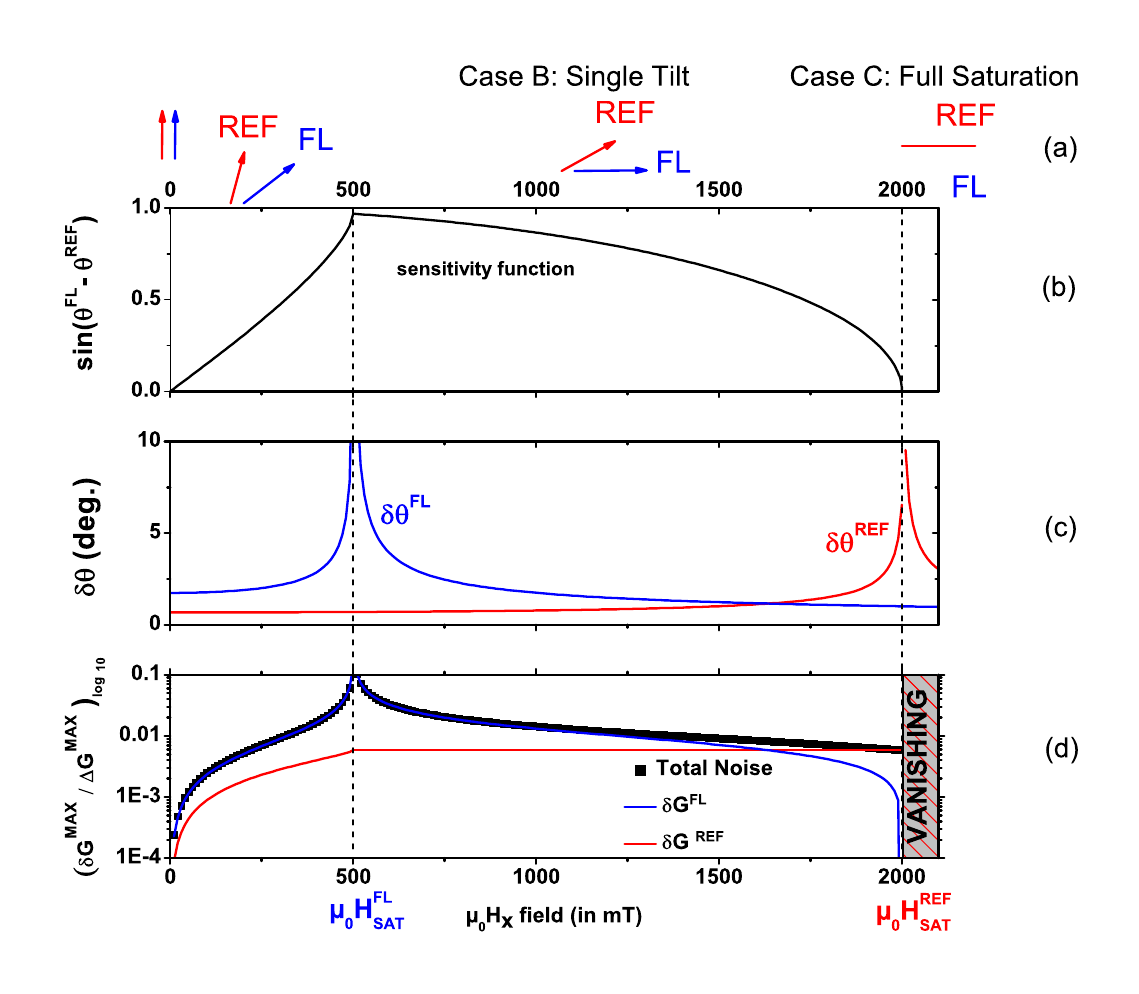}
\caption{Conductance fluctuations (d) are proportional to the product of the sensitivity function (b) by the magnetization fluctuation (c) of the considered layer. a) Sketch of REF and FL magnetizations for the three regimes (i.e. dual tilt, single tilt and dual saturation). b) Sensivity function c) Magnetization fluctuations of FL and REF based on eq.~\ref{DeltaThetaBeforeSaturation} and eq.~\ref{DeltaThetaAfterSaturation} $\textrm{~for}\ H_\textrm {x}<H_ \textrm {sat} $ and $\textrm{~for}\ H_\textrm {x}>H_\textrm {sat} $ respectively d) Order of magnitude of the relative conductance based on eq.~\ref{deltaG_formule_generale}, expected at a temperature of 300K for various fields applied along the in-plane direction. Above $H^{REF}_\textrm {sat}$, there is no conductance fluctuations (shaded area of the log$_{10}$ function).  The paramaters are for a 100\% TMR junction with $R_\textrm {P}$ = 4 k$\Omega$ and $R_\textrm {AP}$ = 8 k$\Omega$, with a junction surface of 50$\times$100 nm$^2$, a FL thickness of 1.8 nm, a magnetization of 1 MA/m and saturation fields values such as $H^\textrm {FL}_\textrm {sat}$=0.5 T and $H^\textrm {REF}_\textrm {sat}$=2T.} 
\end{figure*}


\textbf{Case B:}  At intermediate fields when $H^\textrm {FL}_\textrm {sat} < |H_\textrm {x}| < H^\textrm {sat}_\textrm {REF}$), the REF magnetization is tilted while FL is saturated along $x$. 
The conductance fluctuations are:
\begin{small}
\begin{equation}
 \delta G_\textrm {\textrm{single tilt}}^\textrm {\textrm{FL~fluctuating}} =  \frac{\Delta G_\textrm {max}}{2} \cos \theta^\textrm {REF}_\textrm {0} \beta^\textrm {FL} \sqrt{\frac{1}{H_\textrm {x}-H^\textrm {FL}_\textrm {sat} }} 
 \label{FL_single_conductance}
\end{equation}
 \end{small}

\begin{footnotesize}
\begin{flushright}
with $\cos \theta_0^\textrm {REF} = \sqrt{1- (H_\textrm {x}/H_\textrm {sat, ~REF})^2}$
\end{flushright}
\end{footnotesize}

\textbf{Case C:} At the largest fields when $H^\textrm {FL}_\textrm {sat} < H_\textrm {sat}^\textrm {REF} < |H_\textrm {x}|$ both layers are fully saturated along $x$. In this case, the collinearity of the magnetizations prevents the FL fluctuations to translate into conductance fluctuations: the conductance is constant.

\subsubsection{Conductance noise induced by Reference Layer fluctuations}

In a similar manner, we can estimate the conductance noise induced by reference layer magnetization fluctuations $\delta G_\textrm {\textrm{dual tilt}}^\textrm {\textrm{REF~fluctuating}}$ and $\delta G_\textrm {\textrm{single tilt}}^\textrm {\textrm{REF~fluctuating}}$ in the two field intervals of interest. The formulas are obtained by replacing $H^\textrm {FL}_\textrm {sat}$ by $H_\textrm {sat}^\textrm {REF}$,  $V_\textrm{FL}$ by  $V_\textrm {REF}$ and $M_\textrm{S}^\textrm{FL}$ by $M_\textrm{S}^\textrm{REF}$ under the square root factors in eq.~\ref{FL_dual_conductance} and \ref{FL_single_conductance}.
The most interesting case is case B, where we have :

\begin{equation}
\delta G_\textrm {\textrm{single tilt}}^\textrm {\textrm{REF~fluctuating}} =  \frac{\Delta G_\textrm {max}}{2}  \beta^\textrm {REF} \sqrt{\frac{1}{H_\textrm {sat}^\textrm {REF}}} 
\label{SingleTiltREF}
\end{equation}

\begin{flushright}
where $\beta^\textrm {REF}=\sqrt[]{\frac{\mathrm k_\textrm {B} \mathrm T  }{\mu_\textrm {0} M_\textrm {S}^\textrm {REF} \textrm V_\textrm {\textrm REF} }}$ \\
\end{flushright}

\vspace{0.5cm}
Before saturation of the FL (case A), the conductance magnitude induced by REF magnetization fluctuations is:

\begin{equation}
\delta G_\textrm {\textrm{dual tilt}}^\textrm {\textrm{REF~fluctuating}} =  {\frac{\Delta G_\textrm {max}}{2}}   \beta^\textrm {REF} h^\textrm {REF}_\textrm{Dual} 
\label{DualTiltREF}
\end{equation}

\begin{footnotesize}
\begin{flushright}
with  $h^\textrm {REF}_\textrm{Dual} = \Biggl(\frac{H_\textrm {x}}{H_\textrm {sat}^\textrm {FL}} - \frac{H_\textrm {x}}{H_\textrm {sat}^\textrm {REF}} \sqrt{\frac{1- ({H_\textrm {x}}/{H_\textrm {sat}^\textrm {FL})^2}}{1-({H_\textrm {x}}/{H_\textrm {sat}^\textrm {REF})^2}}}\Biggr)$
\end{flushright}
\end{footnotesize}

\subsubsection{Total conductance noise}

We can finally calculate the standard deviation of conductance $\delta G_\textrm {total}= \sqrt{(
{\delta G^\textrm {\textrm{FL~ fluctuating}}})^2+({\delta G^\textrm {\textrm{REF~ fluctuating}}})^2}$. Equivalently, we can estimate the resistance fluctuations and notice that they increase with the temperature as 
\begin{equation} 
 \delta R_\textrm {total} = R_\textrm{MTJ}^2 \delta G_\textrm {total} \propto \sqrt{T}
 \end{equation}

Fig. 4(d) gathers the normalized conductance fluctuations expected for the junction on which we have done our measurements versus the in-plane field. Note that strictly no signal is expected at zero field (as observed experimentally, see inset in Fig. 3) and above the saturation of the reference layer. In addition the signal is expected to be maximal in the vicinity of the saturation of the Free Layer, also in line with our observations.
 
\subsubsection{Optimal field condition}

From the flatness of the conductance noise related to the REF layer (red curve in Fig. 4[d]), one could think that any field in the single tilt configuration would be equally optimal hence adequate for the estimation of the REF magnetization noise. However in practice all fields within that interval do not yield similar sensitivities and are affected differently by various artefacts. The reason is the following. \\
The noise divergence near the FL saturation induces gain compression of the measurement chain, which prevents faithful measurement near that field of 0.5 T or directly above. At fields higher than 1.5 T, the signal to noise ratio tends to degrade. Besides, the description of the REF magnetization using Eq.~\ref{energy} has proven more accurate near the FL saturation field, because the sensitivity function [Fig.~4(b)] would remain close to unity even if substantial errors would be made on the properties of the REF system. For these reasons, we have thus found that the spectra recorded at fields of 1 and -1 Tesla were optimal, and we shall deduce the device temperature from these data sets.

At this stage, a word of caution is needed. At this specific fields of 1 and -1 T, the FL and REF modes frequencies are close but separated [see Fig.~2(c)]. However, the peaks corresponding to the FMR of the REF mode exhibit a shoulder on the high frequencies side [Fig.~2(c)].  In the small voltage interval where this shoulder is detected unambiguously, its frequency versus field dispersion curve (not shown) is parallel to that of the FL quasi-uniform FMR mode and they share similar linewidths; this argues for the shoulder to arise from a non uniform spin wave mode of the FL. In that case, the energy stored in this fluctuator should not be taken into account when evaluating the power stored in the REF uniform mode. 

Although we believe that the shoulder does correspond to a FL non uniform mode, the voltage and field intervals on which we observe the shoulder is too narrow to draw a definitive conclusion. Therefore in the remainder of the paper, we will evaluate the two hypotheses. In Fig.~5 we will assume that the power contained in the shoulder belongs to the FL and should thus be discarded when analyzing the REF. Conversely in Fig.~6, we will consider the less probable scenario that the power contained in the shoulder belongs to the REF and has to be counted for in the estimation of the REF layer fluctuation amplitude.

\subsection{Thermodynamical estimation of temperature from the voltage noise}

Let us now evaluate how the conductance fluctuations can be used to deduce the device temperature. In the experiments, the measured signal is the voltage noise delivered to the amplifiers of input impedance $Z_\textrm {load} \approx 50~\Omega$. The voltage source is the electromotive force $\delta R \times I_\textrm {dc}$ created by the resistance noise and the dc current, and passing though the internal resistance $R_\textrm {MTJ}$ of the MTJ. Considering that $Z_\textrm {load} << R_\textrm {MTJ}$, the standard deviation $\delta V$ of the voltage noise delivered to the amplifiers can thus be written as:

\begin{equation}
\delta V = \frac{Z_\textrm {load} V_\textrm {dc} \delta R_\textrm {total}}{2 \sqrt{2} (R_\textrm {MTJ})^2} = \frac{Z_\textrm {load}}{2 \sqrt{2}} \times \delta G_\textrm {total} \times V_\textrm {dc}
\label{deltaV_exp}
\end{equation}

The last notation implies that $\delta V ^2  \propto T \times V_\textrm {dc}^2$, such that if the noise power increases faster than $V_\textrm {dc}^2$ this will indicate that the sample is either heated by the current, or that its fluctuations are amplified by the spin-torque.


As seen by the asymmetry in Fig.~3(a), the free layer is subject to a substantial STT affecting its fluctuations that in turn cannot be used to  access the sample temperature. In contrast, the high volume and the high damping of the reference layer makes it essentially insensitive to spin torque, such that the amplitude of its fluctuation should be a reliable measurement of the thermal level. Fortunately, the fluctuations of FL and REF occur at very distinct frequencies (Fig. 3) so that we can integrate their corresponding signals separately.
In practice, we will look at the voltage noise power normalized by the squared voltage, by defining the criterion $Q^\textrm {MAG}$ which also corrects for the voltage dependence of the magneto-resistance ratio $\Delta G_\textrm {max}$ [Fig. 1(d)] :

\begin{equation}
Q^\textrm {MAG} = \frac{1}{V_\textrm {dc}^2} \times \frac{8}{Z_\textrm {load}^2 \Delta G_\textrm {max}^2} \times 
\int \limits_\textrm {mode} {\frac{d (\delta V^2)}{df}df}
\label{Qmag}
\end{equation}

where $\frac{d (\delta V^2)}{df}$ is the spectral density of the total voltage noise delivered to the amplifier and the integration is carried on a given spin wave mode (REF or FL). 
Using eq.~\ref{FL_single_conductance}, \ref{SingleTiltREF} and \ref{deltaV_exp} in case B) when only the FL is saturated (case B) we expect in the absence of spin-torque:

\begin{equation} Q^\textrm {MAG}_\textrm {FL} =   \frac{1- (H_\textrm {x}/H_\textrm {sat}^\textrm {REF})^2}{H_\textrm {x}-H^\textrm {FL}_\textrm {sat}} \times \frac{\mathrm k_\textrm {B} \mathrm T}{4 \mu_\textrm {0} M_\textrm {S}^\textrm {FL} \textrm V_\textrm {\textrm FL}}
\end{equation}

and

\begin{equation}
Q^\textrm {MAG}_\textrm {REF} =  \frac{\mathrm k_\textrm {B} \mathrm T}{4~ \mu_\textrm {0} M_\textrm {S}^\textrm {REF}H_\textrm {sat}^\textrm {REF}~ \textrm V_\textrm {\textrm REF}}
\label{QREFnew}
\end{equation}

The above notation makes it clear that $Q^\textrm {MAG}_\textrm {REF}$ is constant in case B till it abruptly reaches zero in case C. Thanks to the normalization by the bias dependence of the TMR, $Q^\textrm {MAG}_\textrm {REF}$ should not depend on the applied voltage except if the junction is heated by the current, i.e. we expect that $\forall H_\textrm {x} \textrm{, we have }  Q^\textrm {MAG}_\textrm {REF}  \propto T$ and we will use this final result to estimate the current-induced temperature rise (Fig. 5 and 6). \\
Note that the proportionality is only true only provided that the magnetization orientations are not affected by the spin torque (see the trigonometric factors in Eq.~\ref{deltaG_formule_generale}); this fact can be verified by looking at the junction's resistance and the related errors are minimized when the FL is firmly saturated.


\begin{figure}
\includegraphics[width=8.5cm]{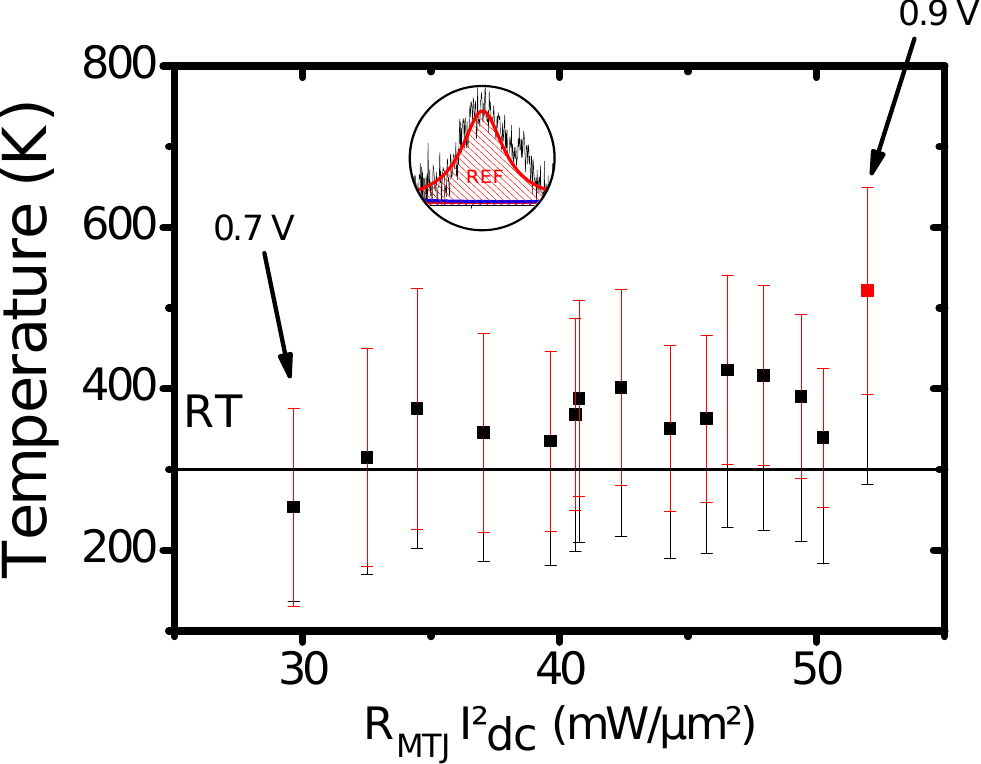}
\caption{Evolution of Temperature in the Reference layer as a function of the Joule power dissipated in the MTJ. For this figure only, we assume that the apparent shoulder in the REF peak [Fig. 2(c)] is a signature of FL fluctuators and should not be taken into account to evaluate the conductance noise coming from the REF layer fluctuations. The red dot stands for 0.9V of bias (T=520K). The back error bars is the confidence interval coming from potential errors in the evaluation of the sensitivity function. The red error bars are the contributions from the Johnson-Nyquist noise.}
\end{figure} 


\begin{figure}
\includegraphics[width=9cm]{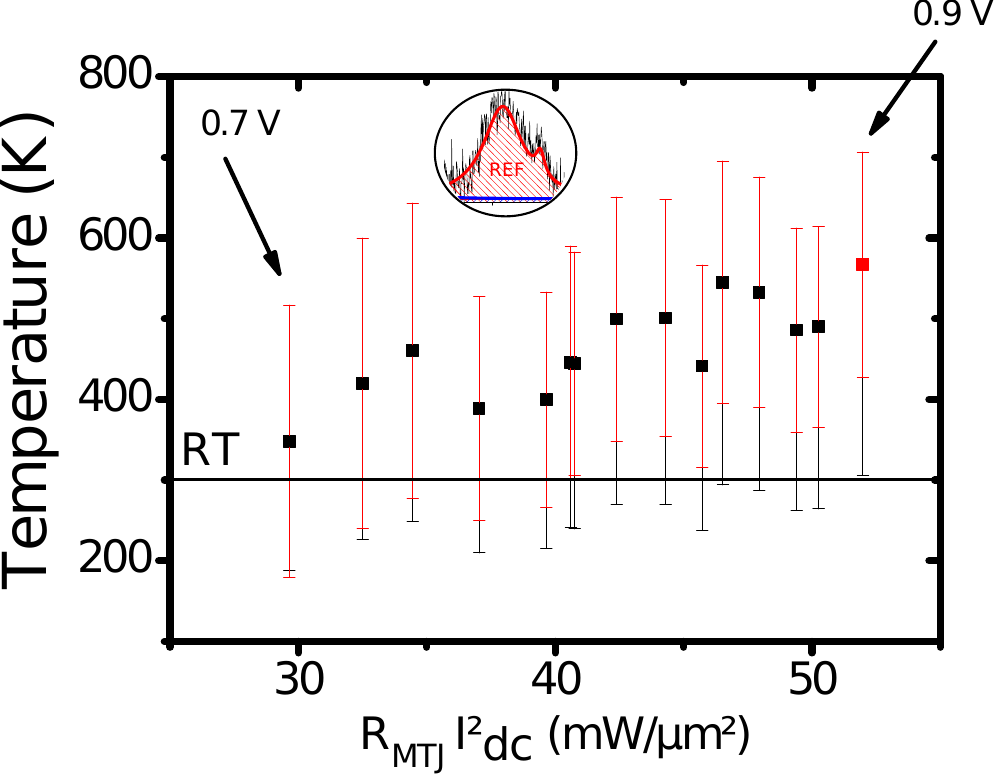}
\caption{Evolution of Temperature in the Reference layer as a function of the Joule power dissipated in the MTJ. For this figure only, we assume that the apparent shoulder in the REF peak [Fig. 2(c)] corresponds to REF layer fluctuations and should be taken into account to evaluate the conductance noise coming from the REF layer fluctuations. The red dot stands for 0.9V of bias (T=570K). The back error bars is the confidence interval coming from potential errors in the evaluation of the sensitivity function. The red error bars are the contributions from the Johnson-Nyquist noise.} 
\end{figure}

\subsection{Confidence interval and error bars on the temperature estimates \label{confidence}}

Despite all the precautions taken, the temperatures that can be deduced from the equality between the measured noise (Eq.~\ref{Qmag}) and the expected noise (Eq.~\ref{QREFnew}) can still be affected by systematic errors.

(i) The first source of systematic error is related to our description of the REF layer hysteresis: the temperature estimate relies on the sensitivity function [Fig. 4(b)] which accounts for the angle between REF and FL magnetization. Considering that Eq. \ref{energy} is a crude description of the REF magnetization behavior, we have used the hard axis loops [Fig. 1(c)] that inform on $\cos (\theta^\textrm {REF}_0-\theta^\textrm {FL}_0)$ to evaluate the sensitivity function  $\sin (\theta^\textrm {REF}_0-\theta^\textrm {FL}_0)$ in an alternative manner. As we do not reach full saturation in hard axis hysteresis loops, there is some uncertainty on the angle $\theta^\textrm {REF}_0-\theta^\textrm {FL}_0$. This way of evaluating the sensitivity yields a sensitivity function that is at most 23\% higher than the result deduced from Eq. \ref{energy}. As the temperature is quadratic with the fluctuations amplitude, an error in the sensitivity function yield a maximum overestimation of the temperature by $\frac{\Delta T^\textrm{SF}}{T}$ = 0.5. The corresponding confidence interval is displayed as the black error bars in Figs. 5 and 6.

(ii) The second source of error has its roots in the (non magnetic) Johnson-Nyquist noise that can only be partially subtracted by our experimental procedure (¤\ref{exp}). Its spectral density is $\frac{d (\delta V^2)}{df}  = 4 ~k_\textrm B T ~Z_\textrm{load} . 10^\frac{NF}{20}$ at the input stage of the amplifier of noise figure $NF$. Depending on the experimental settings, a fraction $\epsilon\approx 0.1$ of this noise power remains and should be plugged in an equation similar to Eq.~\ref{Qmag} to yield the measurement sensitivity floor $Q_\textrm {REF}^\textrm{JN}$ which reads 
\begin{equation}
Q_\textrm {REF}^\textrm{JN} = \frac{\epsilon}{V_\textrm {dc}^2} \frac{4 \mathrm k_\textrm {B} T Z_\mathrm {load}}{Z_\textrm {load}^2 \Delta G_\textrm {max}^2} \times (10^\frac{NF}{20}). \Delta f
\end{equation}
where $\Delta f \approx 2$ GHz is the full width at half maximum of the REF layer mode.
Comparing this floor with Eq.~\ref{QREFnew}, we find that the magnetic noise dominates the remaining Johnson-Nyquist noise if the applied voltage exceeds a detection threshold $V_\textrm{d}$ with
\begin{equation}
V_\textrm{dc}^2 \geq  V_\textrm{d}^2 = 8\epsilon \times 10^\frac{NF}{20} \frac{\mu_0 M_\textrm{S}^\textrm{REF}  H_\textrm{sat}^\textrm{REF} V_{REF}}{\Delta G_\textrm{max}^2 Z_\textrm{load}} \Delta f  
 \label{Vd}
 \end{equation}

With our material parameters, Eq. \ref{Vd} would predict a sensitivity floor of 0.6 V. In practice, spin wave modes could be identified provided of $| V_\textrm{dc}| \geq 0.4~\textrm{V}$ such that we define $V_\textrm{d}$ as 0.4 V. Below this, the Johnson-Nyquist noise exceeds too much the voltage noise delivered by the magnetic fluctuations and the evaluation of the temperature is no longer possible.
The maximum relative error on the temperature due to the remaining part of the Johnson-Nyquist noise is $\frac{\Delta T^\textrm{JN}}{T}= (Q_\textrm {REF}^\textrm{JN}-Q_\textrm {REF}^\textrm{MAG})/Q_\textrm {REF}^\textrm{MAG}$ or equivalently 
\begin{equation}
\frac{\Delta T^\textrm{JN}}{T} = \frac{V_\textrm{d}^2} { V_\textrm{dc}^2 - V_\textrm{d}^2}
\label{JNerrorBars}
\end{equation}

Considering our two sources of potential errors, we could enhance the reliability of our method to estimate the temperature. The uncertainty due to the sensitivity function can be lowered by a more accurate layer description than Eq. \ref{energy} but this would be at the expense of the analytical character of our method. The uncertainty due to the Johnson-Nyquist noise is not of random nature and could be reduced by a factor of almost 2 by using amplifiers with a lower NF around the frequency of interest. Getting lower noise figure would imply to sacrifice the broadband capability of SWNS that is useful to assign each dispersion curve to its hosting magnetic layer. Longer accumulation times could also reduce $\epsilon$. 

\section{Discussion \label{discu}}

The voltage dependence of the sample's absolute temperature (Figs. 5 and 6) can be deduced from the equality between the measure noise of the REF layer (Eq.~\ref{Qmag}) and the expected noise (Eq.~\ref{QREFnew}), with confidence intervals defined using Eq.~\ref{JNerrorBars} and the considerations of ¤\ref{confidence}. The lower bound of the sample temperature is reported in Fig.~5 for which we have assumed that the power contained in the shoulder of the REF peak [see Fig. 2(c)] should be ascribed to non uniform spin waves in free layer and hence be discarded. The upper bound of the sample temperature is given in Fig.~6 when the full high frequency peak is integrated. 
At the highest applied voltages, i.e. when errors are minimal, we find that the injected Joule power of $50 \textrm{~mW}/\mu\textrm{m}^2$ it induces a sizable junction heating of 220 to 270 K. We thus conclude that the heating efficiency is in the range of $4.4 \textrm{~to~} 5.4~ \textrm{K.mW}^{-1}.\mu\textrm{.m}^2$, with more confidence in the lowest value of the heating efficiency. \\ 
We emphasize that our heating efficiency is consistent with anterior independent conclusions that were drawn using alternative methods. H\'erault et al. \cite{herault_nanosecond_2009} used the exchange bias within in-plane magnetizated tunnel junctions to deduce a heating efficiency of $3.5~K \textrm{mW}^{-1}.\mu\textrm{m}^2$. By solving the heat diffusion equations in similar geometries, heating efficiencies of\cite{lee_increase_2008} 1.4 and \cite{hosotani_effect_2010} $5~K \textrm{mW}^{-1}.\mu\textrm{m}^2$ were predicted and seem to corroborate our finding.

In our two temperature estimation scenarios (Figs. 5 and 6)), we conclude that there is essentially no current-induced heating of the MTJ at the voltages of 0.6 V that are needed for the manipulation of the FL magnetization by STT. This conclusion is important in the sense that it proves that it is legitimate to assume a device temperature of about 300K when applying the conventional methods of analysis of the distributions of STT switching currents and switching fields, as performed in many instances when analyzing the thermal stability in STT-MRAM applications \cite{min_study_2010, thomas_perpendicular_2014}.

Conversely at the higher voltage of 0.9~V i.e. when approaching the junction breakdown, the injected Joule power induces a sizable junction heating of 220 to 270 K. This may impact the breakdown behavior and have some consequences for long term device reliability. However in any cases, the MTJ temperature stays much below 750 K which is the temperature during the CMOS back-end-of-line final annealing. We conclude that normal operation should not induce a material fatigue that is more pronounced than the one occurring during the device fabrication.

\section{Conclusion}

We have shown that a direct measurement of the absolute temperature of a nano-sized Magnetic Tunnel Junction (MTJ) can be performed using the high frequency electrical noise that it delivers under a finite voltage bias. The absolute temperature can be deduced with a precision of $\pm 60$~K.  Our method relies on the spectroscopy of the electrical noise that the spin wave population generates though the tunnel magneto-resistance phenomenon in the presence of a biasing current. In practice we perform an analytical treatment of the hard axis field-dependence of MTJ noise spectra. While our method could be used in principle on many types of tunnel junctions, we have illustrated it on perpendicularly magnetized nanopillars that mimic those used in STT-MRAM applications. We find current-induced heating efficiencies of typically 5 Kelvins per $\textrm{mW}/\mu \textrm{m}^2$ which is comparable to the values deduced from other classes of methods.

\section{acknowledgements}

We acknowledge support from Samsung Global MRAM Initiative (SGMI) research program. Useful discussions with Paul Crozat and Joo-Von Kim are also acknowledged.


\bibliography{Publi_JAP_Temperature_ALG}

\end{document}